\documentclass[preprint,showpacs,floats,letterpaper,floatfix,,groupedaddress,eqsecnum]{revtex4}
\usepackage{subfigure}

\usepackage{amssymb,amsmath}
\usepackage[dvips]{graphicx}

\usepackage{dcolumn,epsfig}

\begin{document}

\title{Ratchet effect in an underdamped periodic potential and its characterisation}

\author{Shantu Saikia$^1$}
\email{shantusaikia@anthonys.ac.in}
\affiliation{$^1$St. Anthony's College, Shillong-793001, Meghalaya, India}

\begin{abstract}
 Ratchet effect in a driven underdamped periodic potential system is studied.
The presence of a space dependent and periodic friction coefficient, but with a phase
difference with the symmetric periodic potential is shown to generate substantial 
ratchet current.The ratchet performance is
characterised in terms of the various parameters of transport. The performance 
of this ratchet is compared with a ratchet with an underlying periodic and 
asymmetric potential. It is shown that an optimum combination of inhomogeneity in the
system and asymmetry of the potential can substantially enhance the performance of an 
underdamped ratchet. 
\end{abstract}

\date{\today}
\maketitle

\section{Introduction}
\hspace{0.6cm}   
\par Noise or fluctuations are inherent in almost any physical or biological
system at a finite temperature. These fluctuations play a dominant role in 
the system dynamics at the microscopic domain, particularly when their energy 
scales become comparable to that of the system. 
Different phenomena have been discovered in which these fluctuations play a 
counter intuitively constructive role 
\cite{Julicherc1,Linkec1,Astc1,Mahc1,Reic1, MaddoxAc1, Gammac1}.

\par Ratchet effect \cite{Reic1} is the phenomenon in which directed 
particle transport
takes place in a driven asymmetric periodic potential system without the application
of any obvious external bias, aided by noise or fluctuations. The successful 
application of Ratchet effect in explaining different processes in the 
Biological and Physical world has led to a spurt of studies - both theoretical 
and experimental 
\cite{AstumianAc1,BartuAc1,MagnascoAc1,Rousseletc1,DoeringAc1,Millonasc1}.  
Though primarily studies on Ratchets were aimed at explaining the working
of biological or molecular motors \cite{Spudichc1}, presently it has 
diversified to various other fields \cite{HanggiAc1, Siwyc1, 
Savelevc1, HanggiRevc1}.

\par For realising ratchet effect, the symmetry of the system has to be broken.
Also the system has to be driven away from equilibrium. Based on these conditions
different models of Ratchets have been proposed and studied \cite{Reic1}. 

\par In this work we study ratchet effect in a model {henceforth called as 
\it{inhomogeneous ratchet}}.
In this ratchet, particle transport occurs in a driven underdamped inhomogeneous
 periodic potential system. The asymmetry in the system is due  to the space
dependent friction coefficient which is periodic with the same periodicity as the
potential, but with a phase difference with the potential 
\cite{Jayanc1,DanAc1,MangalBc1,Raishmac1,MangalCc1, 
Saikia1, Saikia2, Wanda1, Wanda2, Wanda3, Don, Saikia4, Saikia3}.

\par Particle transport in inhomogeneous systems driven away from
equilibrium has been an active field of research in the past few
 decades \cite{Pareek,Jayanc1,DanAc1,
MangalBc1,Raishmac1,MangalCc1, Saikia1, Saikia2, Wanda1,Wanda2, Wanda3, Don,
 Saikia4,Saikia3, LandauerAc1,ButtikerAc1,BlantAc1,BenzAc1,LuchsingerAc1}.
In such inhomogeneous systems, the particle experiences a non uniform diffusion.
This may be either  due to a space dependent 
friction \cite{Jayanc1,DanAc1,
MangalBc1, Raishmac1,MangalCc1, Saikia1, Saikia2, Wanda1, Wanda2, Wanda3, Don,
 Saikia4, Saikia3} 
or due to a space dependent temperature in 
the system \cite{LandauerAc1,ButtikerAc1,BlantAc1,BenzAc1,LuchsingerAc1}. 
\par In nature there are numerous examples of systems having 
a space dependent friction. For example molecular motor proteins experiences 
a  space dependent friction coefficient when they move along the periodic
structures of microtubules \cite{LuchsingerAc1}. Brownian motion in confined 
geometries show a space dependent friction \cite{Fucheuxc1}. Particles 
undergoing surface diffusion also experiences a space dependent friction coefficient 
\cite{Fucheuxc1, Brennerc1}. Space dependent friction coefficient also finds
applications in superlattice structures, Josephson Junction equation \cite{FalcoAc1},
and ad-atom motion on the surface of a crystal of identical atoms \cite{Wahnstromc1}. 

\par The various aspects of particle transport in overdamped inhomogeneous
systems with a space dependent friction coefficient has been extensively
studied \cite{Jayanc1,DanAc1,MangalBc1,Raishmac1,DanN}. It was
shown that it is possible to obtain directed particle transport aided
by noise even with a perfectly symmetric and periodic potential in the 
presence of a similarly periodic friction coefficient, but with a phase
difference $\phi$ $(\phi\ne 0,\pi)$ with the potential; the system 
being driven away from equilibrium. The various parameters characterising 
the transport and particle current, like the particle velocity, efficiency
and coherence of transport, diffusion etc. have been studied. Also it has
been shown that forced inhomogeneous ratchets exhibit current reversals.

\par Though overdamped approximation is valid for many physical and
biological processes and systems \cite{Reic1} - for example molecular
motor movement in the Brownian regime - inertial effects can, however,
play an important role in many other situations \cite{MangalCc1,
Saikia1, Saikia2, Wanda1, Wanda2, Wanda3, Don, Saikia4, Saikia3,LandauerAc1,
ButtikerAc1,BlantAc1,BenzAc1,Wahnstromc1,Flac1,Junc1,MachuraAc1,
Katjac1,ConstanAc1,Lin,PReimann2c1}. 
Examples being the diffusion of ad-atoms on a 
crystal surface \cite{Pollakc1}, dissipation in threshold devices 
\cite{BorroAc1}, dislocation of defects in metals \cite{Braunc1} and Josephson
junctions \cite{Benc1}.

\par Particle dynamics in overdamped and underdamped systems are
fundamentally different. In the overdamped regime, inertial effects can be 
ignored. The particle in such a system exhibits a hopping motion
between the wells of the potential. The particles
gets trapped in a potential minima. With noise, the particles
escape from the locked state in a potential well only to be trapped in 
another adjacent well. On the other hand in the underdamped dynamics, 
in a periodic potential, the particle can be in two states of dynamics
- the locked state in which the particles are trapped in a potential
minima and the running state in which because of the particle's momentum
the keep travelling over many periods of the potential. 
In the presence of finite noise the particle undergoes transitions 
between these two dynamical states \cite{Risken}.

\par Underdamped dynamics exhibits some interesting phenomena which
are otherwise not observed in overdamped systems. For example
underdamped systems exhibits a giant enhancement of diffusion
coefficient \cite{Benjamin}. The diffusion coefficient is dependent
on the jump rates out of a potential well and also on the particle
jump lengths. These parameters are in turn dependent on the amount
of damping. In the underdamped regime, the particle dynamics is 
dominated by long jumps \cite{Braun}.

\par Inertial inhomogeneous system with a space dependent temperature has
 been shown to yield directed transport of particles in the absence of 
any external bias \cite{LandauerAc1,ButtikerAc1,BlantAc1,BenzAc1,PReimann2c1} . 

\par Particle transport in driven inertial inhomogeneous system with a 
space dependent friction coefficient has been shown to exhibit other interesting
phenomena also \cite{Saikia1, Saikia2, Wanda1, Wanda2, Wanda3, Don, 
Saikia4, Saikia3}. Particle 
current is observed with associated multiple current reversals in the deterministic
regime \cite{Saikia1}.
When driven with a square drive, apart from substantial particle current in 
optimised parameter regimes, the system also shows dispersionless
particle motion at intermediate time regimes \cite{Saikia2,Saikia4}.
Using this model system the phenomenon of Stochastic Resonance in a 
periodic potential system was obtained 
\cite{Saikia1, Saikia2, Wanda1, Saikia4, Saikia3}. This phenomenon in which a
periodically driven system shows an enhanced response to the external 
drive in the presence of noise or fluctuations, was earlier mostly studied
in bi stable systems. 

\par We further explore different characteristics of particle transport
in a driven underdamped periodic potential system using the same model as in 
\cite{Saikia1, Saikia2, Wanda1, Wanda2, Wanda3, Don, Saikia4, Saikia3}. 
The performance of this {\it{inhomogeneous ratchet}} is compared
with a more commonly studied  model in which 
asymmetry in the system is due to an inherent asymmetry in the underlying
periodic potential (henceforth called as {\it{asymmetric ratchet}})
 \cite{Reic1,Borr, Machu, Jung, Marc}.
A scheme for obtaining enhanced efficiency of a ratchet in the underdamped
regime is proposed by combining essential features of both these ratchets.

\par For optimising the working of ratchets, their performance needs to be 
characterised. 
 As Ratchets operate in environments
dominated by the random fluctuations, their performances are 
measured by the fluctuating quantities \cite{LinkeAc1} like position ($x(t)$),
 velocity ($v$) or work output ($W$).  
In the presence of fluctuations, the particle positions at any given time get
 spread out when being transported as an ensemble through a distance from a common 
initial position and time. A measure of this spread is given
by the effective diffusion or dispersion 
$D_{eff}=\frac{\langle x^2(t) \rangle-\langle x(t)\rangle ^2}{2t}$,
 where the average $<..>$ is over
an ensemble. But as $D_{eff}$ does not distinguish 
between motors of different average velocities, we need a coherency
 parameter which incorporates both velocity and spread. This is given
by the Peclet number \cite{Landauc1} 
$P_e=\frac{\langle v\rangle l}{D_{eff}}$, where $l$ is a characteristic
length of the system and $<v>$ is the time averaged velocity.  
\par Apart from the Peclet number another important parameter for the
 characterisation of a Brownian motor is its efficiency.  
As these motors operate in a viscous environment, they can do work
 against an external load $F$, and against the viscous drag in the medium or both.
 The {\it efficiency of energy conversion} of a motor with respect to the
work done against a load $F$ \cite{Reic1, Sekimotoc1} is given as
$\eta_E=\frac{|F\langle v \rangle|}{P_{in}}$, where $P_{in}$ is the
input power. Though this definition  is thermodynamically correct, it
 becomes unsatisfactory in the case of Brownian motors which works only via the
 transport against the viscous  drag of the medium, in the absence of a load.
  A definition of efficiency which 
takes into account the work done solely against the viscous drag is called the 
{\it Stokes efficiency}, 
$\eta_{S}=\frac{\langle v \rangle ^2}{|\langle v^2 \rangle - T|}$ 
\cite{MachuraAc1}. Suzuki and Munakata \cite{Suzukic1} combined
 the above two efficiencies to give a more general definition of
 efficiency called the rectification efficiency given as
$\eta_{rec}=\frac{\gamma \langle v \rangle ^2 + F\langle v \rangle}{P_{in}}
$.

\section{The model}
In this work we consider the motion of a particle in a 
periodic potential $V(x)=-V_0( \sin(kx)+ b\sin 2kx)$. The friction coefficient 
$\gamma (x)=\gamma_0(1-\lambda \sin(kx+\phi))$ is periodic with the same 
periodicity as the potential but has a phase difference $\phi$. $b$ is the parameter
which determines the underlying asymmetry in the potential whereas $\lambda$ is the
inhomogeneity parameter. The system is studied under the following approximations. 

(i) When the system is inhomogeneous, with a space dependent friction coefficient
but with a symmetric potential. For this, $b = 0$ and $\lambda \ne 0$. In this case 
the asymmetry in the system is only due to the inhomogeneity 
({\it {inhomogeneous ratchet}}).
(ii) When the system is homogeneous but there is an asymmetry in the underlying
periodic potential. For this $b \ne 0$ and $\lambda = 0$ ({\it{asymmetric ratchet}})
(iii) When the system has both inhomogeneity and asymmetry in the potential. In this 
case $b$ and $\lambda$  both are $\ne 0$ and the asymmetry
in the system is due to both the parameters (combination of {\it{inhomogeneous}}
and {\it{asymmetric}} ratchets).

\par The system is driven periodically by an external periodic forcing 
$F(t)$=$F_0$ $\cos(\omega t + \phi_0)$. Over a period of the drive, 
the net external forcing averages out to zero. Hence there is no obvious applied bias
in the system. The system is studied in the presence of finite noise. 

\par The one dimensional equation of motion of a particle of mass $m$ is given by
the Langevin's equation,
\begin{equation}
m\frac{d^{2}x}{dt^{2}}=-\gamma (x)\frac{dx}{dt}-\frac{\partial{V(x)}}{\partial
x}+F(t)+\sqrt{\gamma(x)T}\xi(t).
\end{equation}
In Eq. 1, the temperature $T$ is in units of the Boltzmann constant $k_B$.
The Gaussian distributed fluctuating forces $\xi (t)$ obeys the statistics:
 $<\xi (t)>=0$, and $<\xi(t)\xi(t^{'})>=2\delta(t-t^{'})$. 
The above equation is written in dimensionless units by setting $m=1$, $V_0=0$
and $k=1$. The dimensionless form
of the equation keeping the same symbols for the reduced variables can be 
written as 
\begin{equation}
\frac{d^{2}x}{dt^{2}}=-\gamma(x)\frac{dx}{dt}
+\cos x +2b\cos 2x +F(t)+\sqrt{\gamma(x) T}\xi(t),
\end{equation}
where $\gamma(x)=\gamma_0(1-\lambda \sin(x+\phi))$.
In the reduced equation too, the noise variable, $\xi$, satisfies exactly similar 
statistics as earlier. The 
potential $V(x)$ and the friction coefficient $\gamma(x)$ are periodic having the 
same periodicity of $2\pi$. $F_0$ is amplitude of the external drive
and the particles experience a finite barrier between two consecutive
wells of the potential $V(x)$ for all values of amplitude $F_0$, ($0<F_0<1$).
The barrier disappears momentarily in a period at the critical field value $F_0=F_c=1$.

\par For obtaining finite particle current in a driven periodic potential system, 
the asymmetry of the system needs to be broken. This can be achieved by making the
potential asymmetric ($b \ne 0$) ({\it{asymmetric ratchet}}) or by making the 
system inhomogeneous ($\lambda \ne 0$, with a value of phase difference 
$\phi\neq n\pi$)({\it{inhomogeneous ratchet}}). In this work,
as discussed above we study the system under different approximations.

\section{Numerical results} 
Eq. 2 is solved numerically using the Heun's method \cite{Manne}
 for solving stochastic 
differential equations. The particle trajectories are obtained for given initial 
conditions for fixed values of the parameters. 
The steady state mean velocity
$\langle \bar v \rangle $ of the particle is obtained as
\begin{equation}
 {\langle \bar v \rangle}=   \langle \lim_{t \rightarrow \infty}\frac{x(t)}{t}\rangle,
\end{equation}
The system is allowed to evolve over effective times of around $t=10^5$ and 
ensemble averaging $\langle ... \rangle$ is done over an ensemble of 50 
different initial conditions.

\par We study particle
transport in the model {\it{inhomogeneous ratchet}}. The performance of this 
ratchet is compared with that of the {\it{asymmetric ratchet}}.
Both these ratchets schemes are then combined to evolve
a criterion for enhancing the performance of underdamped ratchets.
The main results of our work are highlighted below.
We also study, the performance of the underdamped ratchets as
a function of the other parameters so as to optimise its performance.
\begin{figure}[htp]
  \centering
  \subfigure[]{\includegraphics[width=6cm,height=6.7cm,angle=-90]{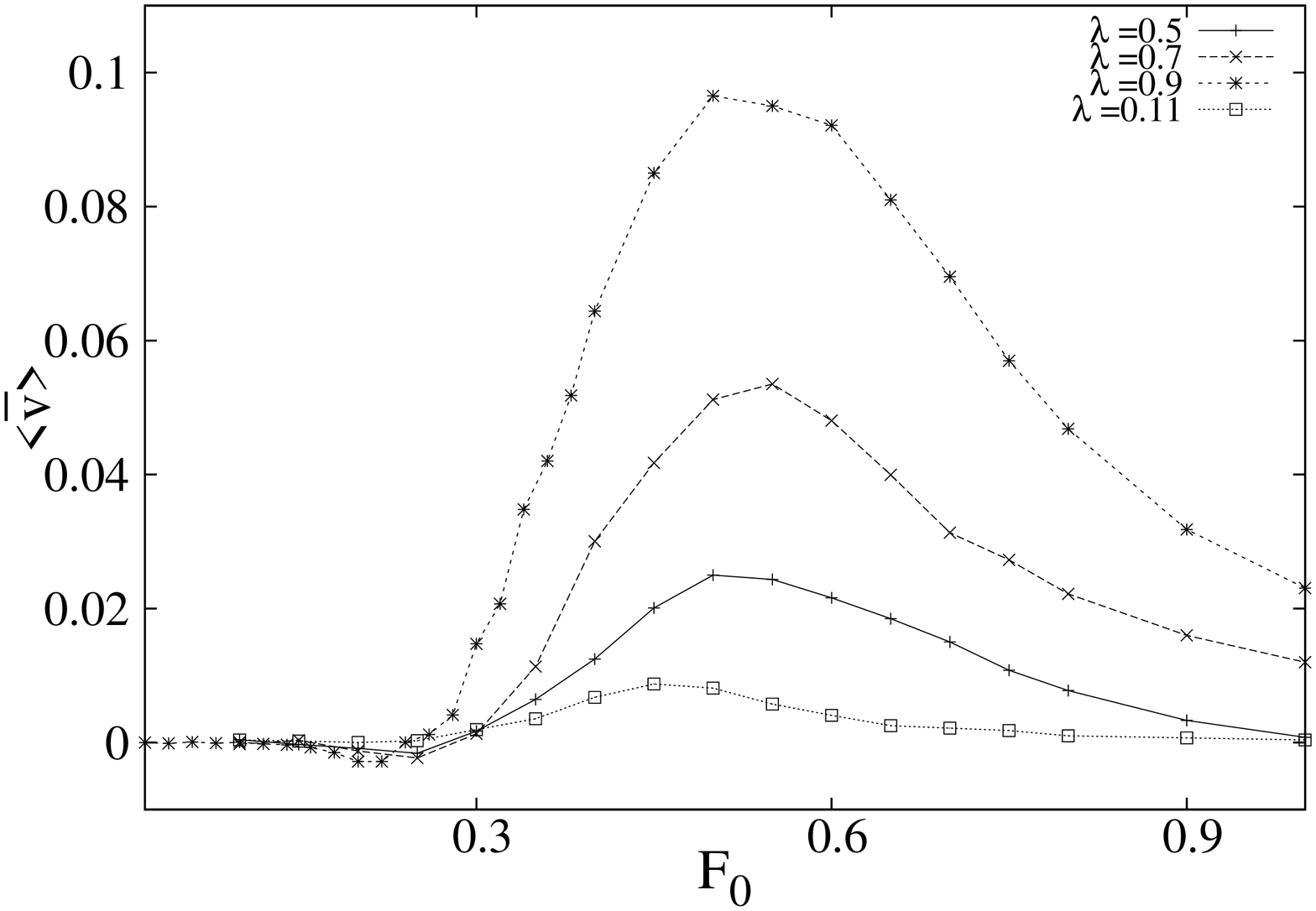}}
\hspace{0.01cm}
\subfigure[]{\includegraphics[width=6cm,height=6.7cm,angle=-90]{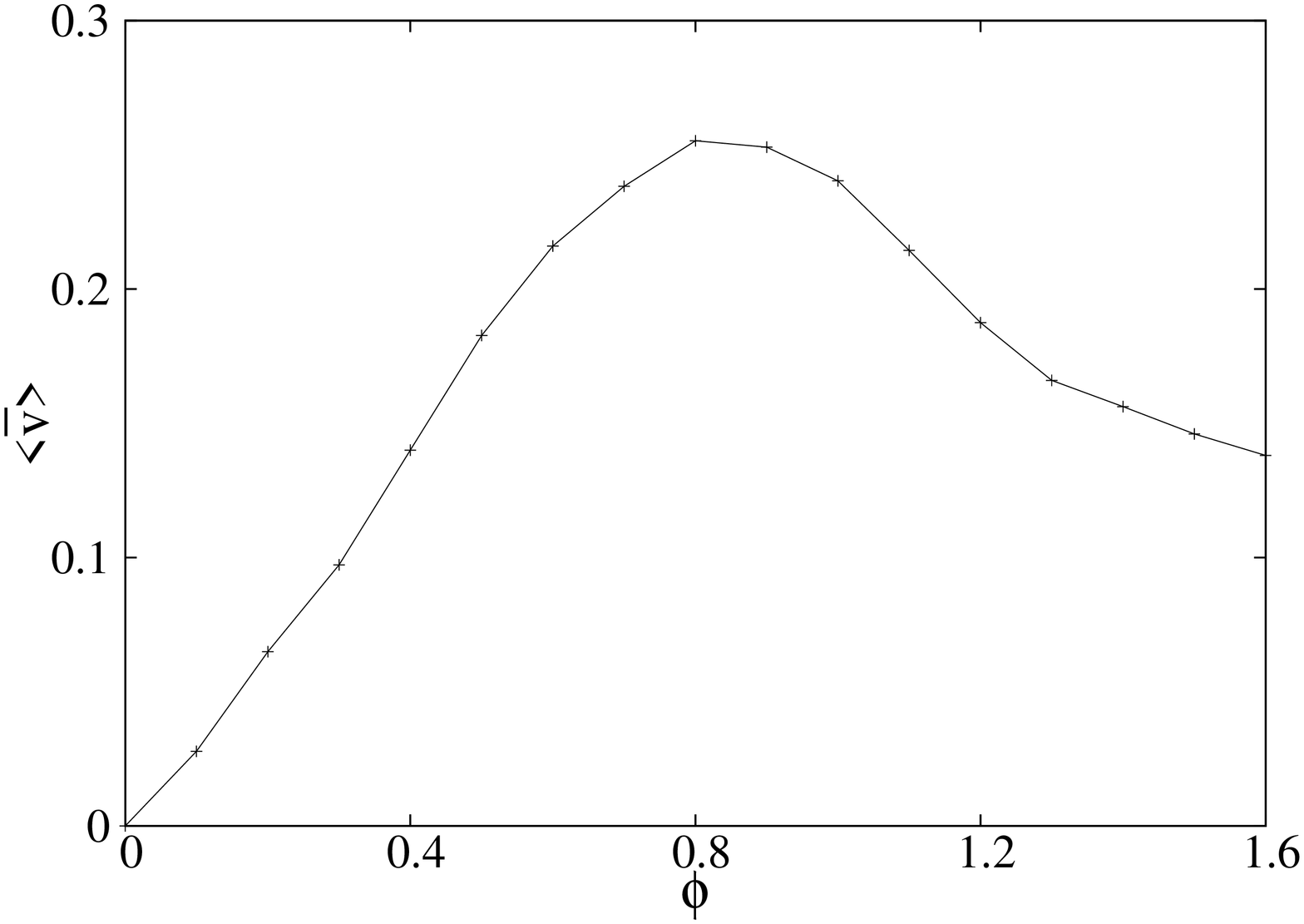}}

\caption{Plot of $\langle \bar v \rangle$ versus $F_0$ for different values of 
$\lambda$ (a) for $\gamma_0=0.12$, $\tau =1000$, $\phi=0.35$ and
$T=0.4$. Plot (b) shows the variation of  $\langle \bar v \rangle$ with $\phi$
for $\lambda = 0.9$, $F_0 = 0.5$ and other parameters remaining same as plot (a).}
\label{fig:edge}
\end{figure}

 \par In Fig.1a the value of average velocity $\langle \bar v \rangle$ is 
plotted as a function of the driving amplitude $F_0$ for different values of 
$\lambda$, keeping the potential symmetric (i.e. $b=0$). With 
a finite value of the parameter $\lambda$ (for a fixed value of $\phi$), 
appreciable ratchet current is obtained. This current is solely due to the
asymmetry introduced into the system because of the space dependent friction
coefficient. The amount of current in this {\it{inhomogeneous ratchet}} however
expectedly depends on the value of the inhomogeneity parameter $\lambda$ 
(for a fixed value of $\phi$).

\par The nature of variation of the particle current in Fig.1a can be understood
by looking at the origin of particle current. 
The magnitude and direction of particle current - if any - in a driven periodic 
potential depends on the number of interwell particle hoppings to the right 
and the left
of a potential minima. The number of such hops are intricately dependent on the
temperature, asymmetry and amplitude of external drive. If there is no asymetry 
in the system, the hoppings to the right and the left are equally probable
giving rise to zero particle current.  
\par In the model inhomogeneous system under our consideration, the asymmetry in
 the system arises because of the phase difference $\phi$ between 
the periodic potential and the space dependent friction coefficient. When $\phi$
is equal to $0$, $\pi$ and $2\pi$, the friction coefficient about the potential 
peak is symmetric. However, for $\phi=0.35$ as considered here, the friction 
coefficient on the left of the potential peaks is larger than on the right.
If this system is driven adiabatically \cite{Wanda2} or with a finite but small
frequency\cite{Saikia2,Wanda3}, this leads to an asymmetric 
particle distribution about the bottom of the potential
well in the dynamic situation, leading to a probability of getting a negative
particle current. Also, because of higher friction, the particles will spend
more time on the left side of the potential peak, absorbing more heat from the
fluctuations
leading to higher temperature. This gives rise to a probability of getting a 
particle current in the positive direction. The final direction of particle current
that is obtained is a result of these two competing effects and also 
is determined by the other parameters of the system. 
\par However, in a recent work \cite{Don} it was shown that if the frequency
of the external drive is large, at low temperatures the particles spend equal
time on either side of the potential peak irrespective of the friction 
coefficient and the motion remains symmetrical in both directions. However as 
temperature rises, for finite $\lambda$ difference starts appearing leading to 
more probability of the particle crossing over to one side from a well than 
the other.

\par Therefore in the inhomogeneous system under our consideration, the final 
magnitude and direction of current is a result of a complex interplay
between the non linearity of the system, the particle dynamics, the asymmetry 
of the system and the choice of the parameters characterising the system.

\par When the amplitude of the external drive is small, the particles remain near the 
bottom of the potential wells. So they do not feel the effect of asymmetry that 
much and they make occasional hoppings out of the well aided by the fluctuations. 
Hence the particle current is less. As the amplitude increases, the particles 
can climb further up the potential walls and they experience the asymmetry in 
thes system more. Also the probability of interwell hoppings aided by fluctuations 
increases. So the current rises. For very high amplitude though, the barrier to
 motion is almost disappears – it disappears momemtarily in each drive if the 
amplitude is more than the critical tilt. So the effect of asymmetry is less 
pronounced leading to less current. In between these two extremes,
 expectedly the current peaks for an intermediate value of $F_0$ (Fig. 1a). 
The amount of current of course
depends on the degree of inhomogeniety in the system which is determined by the 
parameter $\lambda$. It is observed that as the value of the parameter $\lambda$
increases, the particle current also increases monotonically (Fig.1a). 

\par The degree of asymmetry in this system is intricately linked with the 
phase difference $\phi$ between the periodic friction coefficient and the underlying 
periodic potential. Fig. 1b, shows the plot of $\langle \bar v \rangle$ as a 
function of $\phi$. Other parameters remaining fixed, the ratchet 
current is seen to peak for a value of $\phi = 0.8$.

\par This observed variation can be understood by noting that the asymmetry of
the system changes with the value of the parameter $\phi$. For $\phi=0, \pi$ 
and $2\pi$, the friction coefficient is symmetric
about a potential minima leading to zero particle current. For other values of
$\phi$ a finite current can be observed as it leads to a difference in the
number of particle hoppings to the right and left of a potential minima. 
 Asymmetry in the system is maximum for $\phi=0.5\pi$. However maximum 
asymmetry in the system does not necessarily mean the maximum difference in 
the number of hoppings to the left and the right of a potential well, as this 
is also dependent on the temperature of the system \cite{Don}.
Other parameters remaining fixed, current peaks for that value of $\phi$ at which 
this difference is maximum. 
\par In a recent work \cite{Don} a detailed explanation of this has 
been put forward. The average friction coefficient experienced by the particle
and hence its mobility is dependent on $\phi$, which along with the temperature 
of the system determines the asymmetry in the hoppings.

\begin{figure}[htp]
  \centering
\includegraphics[width=6cm,height=6.7cm,angle=-90]{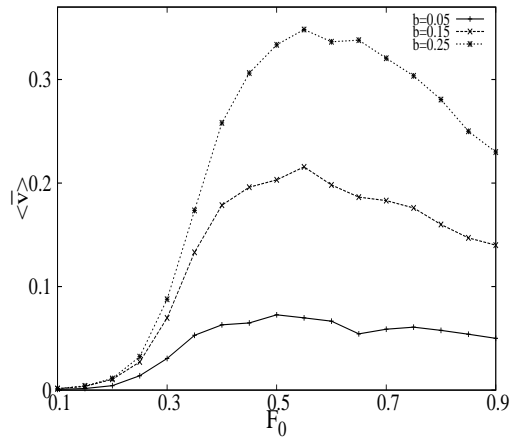}
\caption{Plot of $\langle \bar v \rangle$ versus $F_0$ for different values of 
$b$, for $\gamma_0=0.12$, $\tau =1000$, $\phi=0.35$ and $T=0.4$.} 
\label{fig:edge}
\end{figure}

\par Fig. 2 is a plot of the average particle velocity $\langle \bar v \rangle$ 
for the {\it{asymmetric ratchet}} in which the particle experiences a homogeneous 
friction coefficient ($\lambda = 0$).
The asymmetry in the system is due to the asymmetry in the underlying 
periodic potential.
In the figure the average velocity is plotted as a function of the driving 
amplitude $F_0$ for different values of the parameter $b$ which denotes different 
degrees of asymmetry of the potential. Undoubtedly, higher degrees of asymmetry
in the potential gives rise to higher values of particle current. 
\par The observed variation of the particle current with the amplitude
for the {\it{asymmetric ratchet}} has the same explanation as that 
for the inhomogeneous ratchet as given above, with the current peaking
at an optimal drive amplitude. 

\par Comparison of Fig. 1a and Fig. 2 shows that that in the 
underdamped regime, an {\it{asymmetric ratchet}} (Fig. 2) gives higher values of
particle current than the {\it{inhomogeneous ratchet}} (Fig. 1a) as
inhomogeneity is a feeble cause of asymmetry as compared to the inherent
asymmetry in the potential itself.
However for certain parameter regimes, the {\it{inhomogeneous ratchet}} 
does give currents higher
than that for the {\it{asymmetric ratchet}}. For example current for the 
{\it inhomogeneous ratchet} for $\lambda=0.9$ is higher than that 
for the {\it asymmetric ratchet} for $b=0.05$
Thus, though inhomogeneity is a feeble cause of asymmetry, appreciable 
ratchet currents can be obtained. As particle transport in such 
inhomogeneous system finds analogies in various physical and biological systems as 
mentioned in earlier sections, the {\it inhomogeneous ratchet} provides an 
interesting option for modelling and studying different natural systems.

\par The drive amplitude at which the particle current peaks will also be
dependent on the temperature of the system. For the present case when the
drive frequency is very small compared to the natural frequency of the 
particles at the bottom of a potential minima, in the absence of any fluctuations
($T=0$), the particles remain near the bottom of the potential minima. 
No particle current will be observed in this case. However for finite 
temperature the particle gets kicked around by the fluctuations and they
start exploring the asymmetry of the system. For lower temperatures,
if the drive amplitude is small, the probability of the particles overcoming
the barrier and hopping to another well is less. However as the amplitude 
increases the particle moves up further away from the bottom of the well and
aided by the fluctuations cross over to another well. The asymmetry in the 
system will give rise to preferentially more hoppings in a particular direction
leading to finite current. However, if the temperature of the system is higher
the particles will undergo interwell hoppings at a lesser amplitude of drive.
So it can be concluded that the drive amplitude at which the current peaks
shifts to higher values for lower temperatures.

\section{Characterisation of the Ratchet current}
For optimisation of the performance of a ratchet it needs to be characterised
in terms of the various parameters which defines the quality of transport. One
such parameter is the diffusion coefficient $D_{eff}$. As the
ratchets operate in a very noisy environment which is dominated by
random fluctuations, the particles in a ratchet spreads during its motion over
a period of time. So
apart from calculating the average particle velocity, it is necessary to have
a measure of the diffusive spread of the particles. The measure of this spread
is given by the diffusion coefficient
$D_{eff} = (\langle x^2(t) \rangle - \langle x(t) \rangle^2)/2t$, where
the averaging $\langle ..... \rangle$ is over the ensembles.

\begin{figure}[htp]
  \centering
  \subfigure[]{\includegraphics[width=6cm,height=6.7cm,angle=-90]{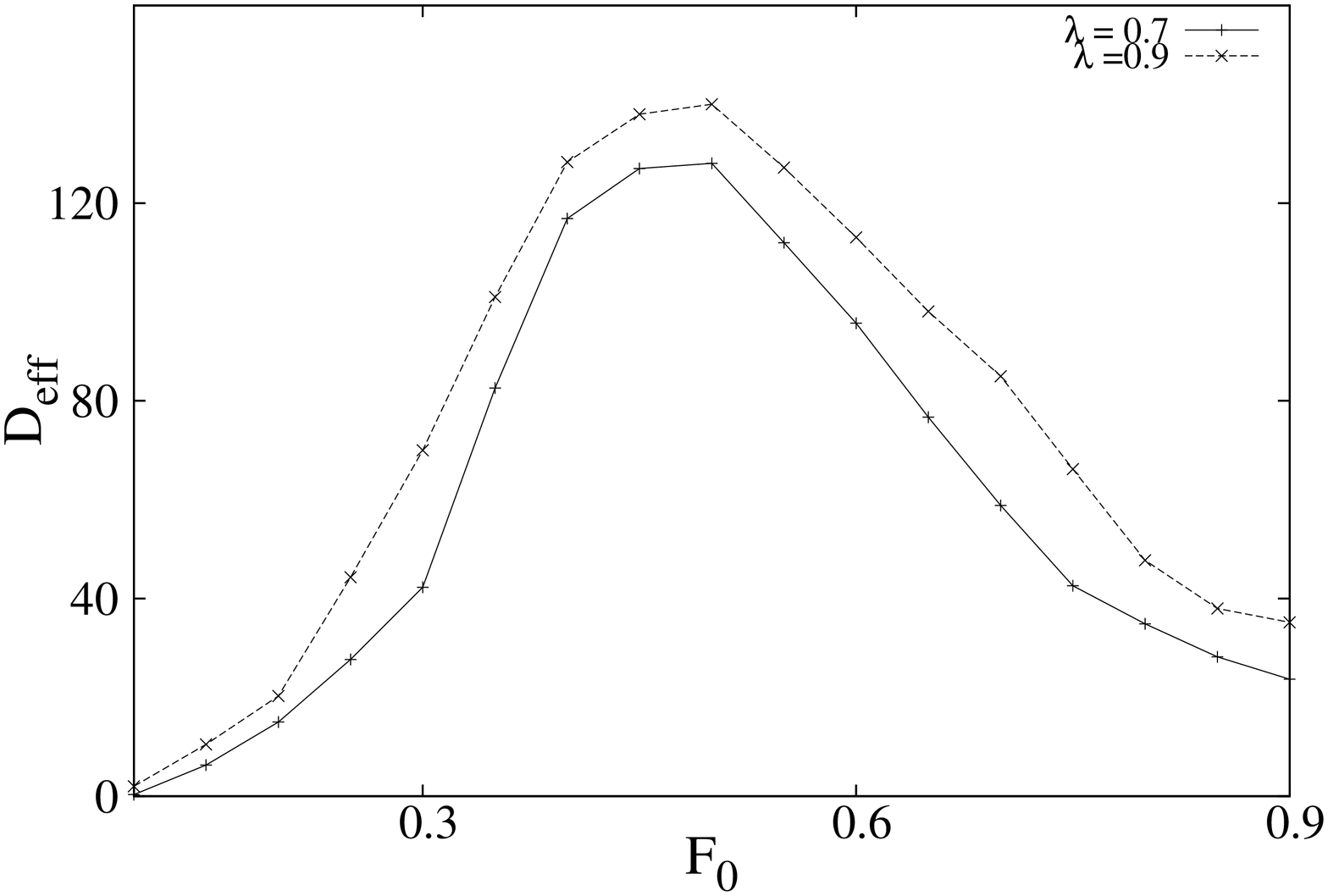}}
\hspace{0.01cm}
\subfigure[]{\includegraphics[width=6cm,height=6.7cm,angle=-90]{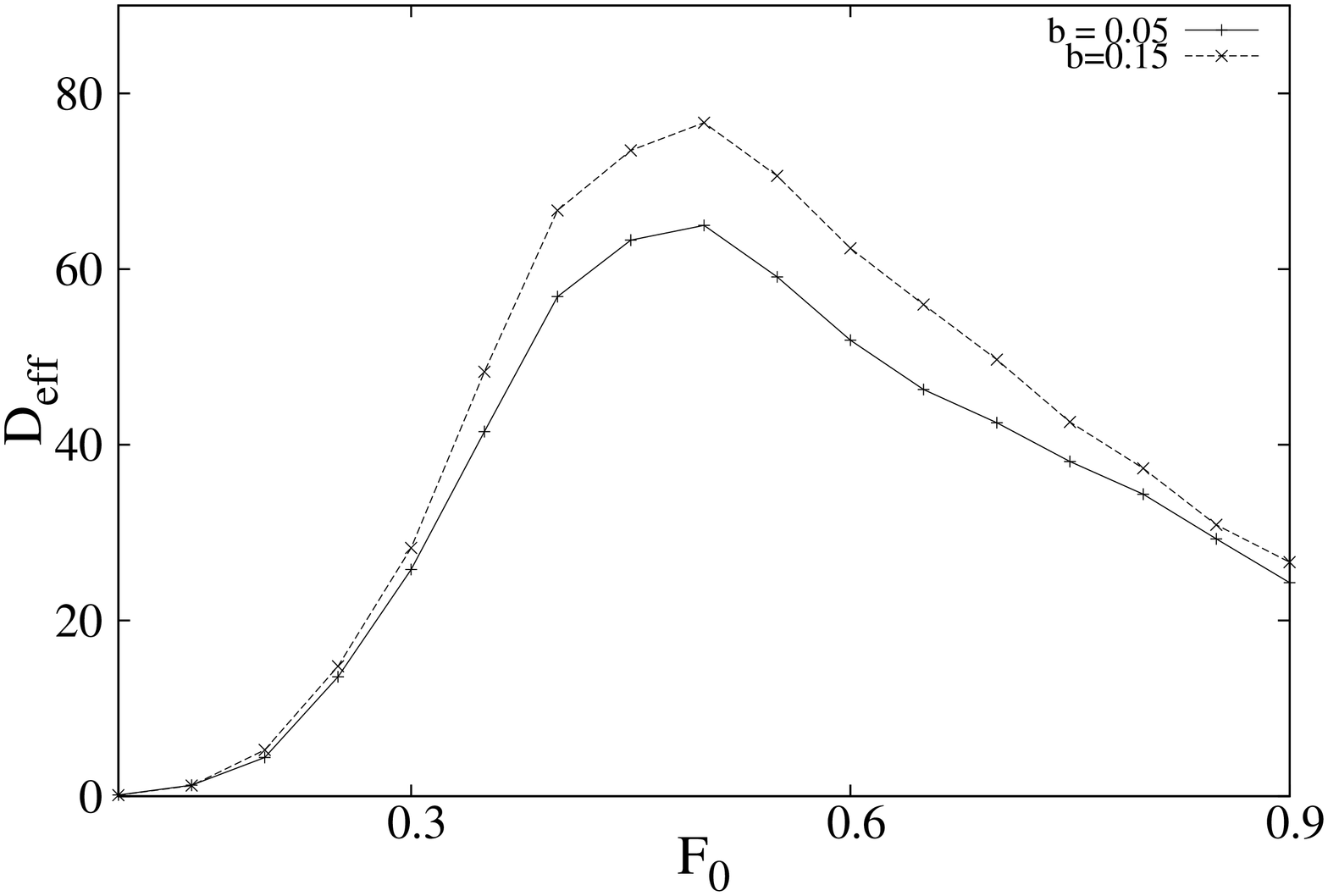}}

\caption{Plot of diffusion coefficient $D_{eff}$ versus $F_0$ for different values of 
$\lambda$ (a) and $b$ (b) for $\gamma_0=0.12$, $\tau =1000$, $\phi=0.35$ and
$T=0.4$.} 
\label{fig:edge}
\end{figure}

\par Fig. 3a and 3b shows the plot of the diffusion coefficient $D_{eff}$
 versus the driving amplitude for the {\it{inhomogeneous ratchet}} and the 
{\it{asymmetric ratchet}}.
Diffusion in both the ratchets peaks around the value of 
$F_0$ at which there is a peak in the average velocity. The degree of spread
in the {\it{inhomogeneous ratchet}} is comparatively much higher. This is
beacause the degree of asymmetry in the system experienced by the particle
leading to its directed motion is higher in the {\it asymmetric ratchet} 
compared to the {\it inhomogeneous ratchet}

\begin{figure}[htp]
  \centering
  \subfigure[]{\includegraphics[width=6cm,height=6.7cm,angle=-90]{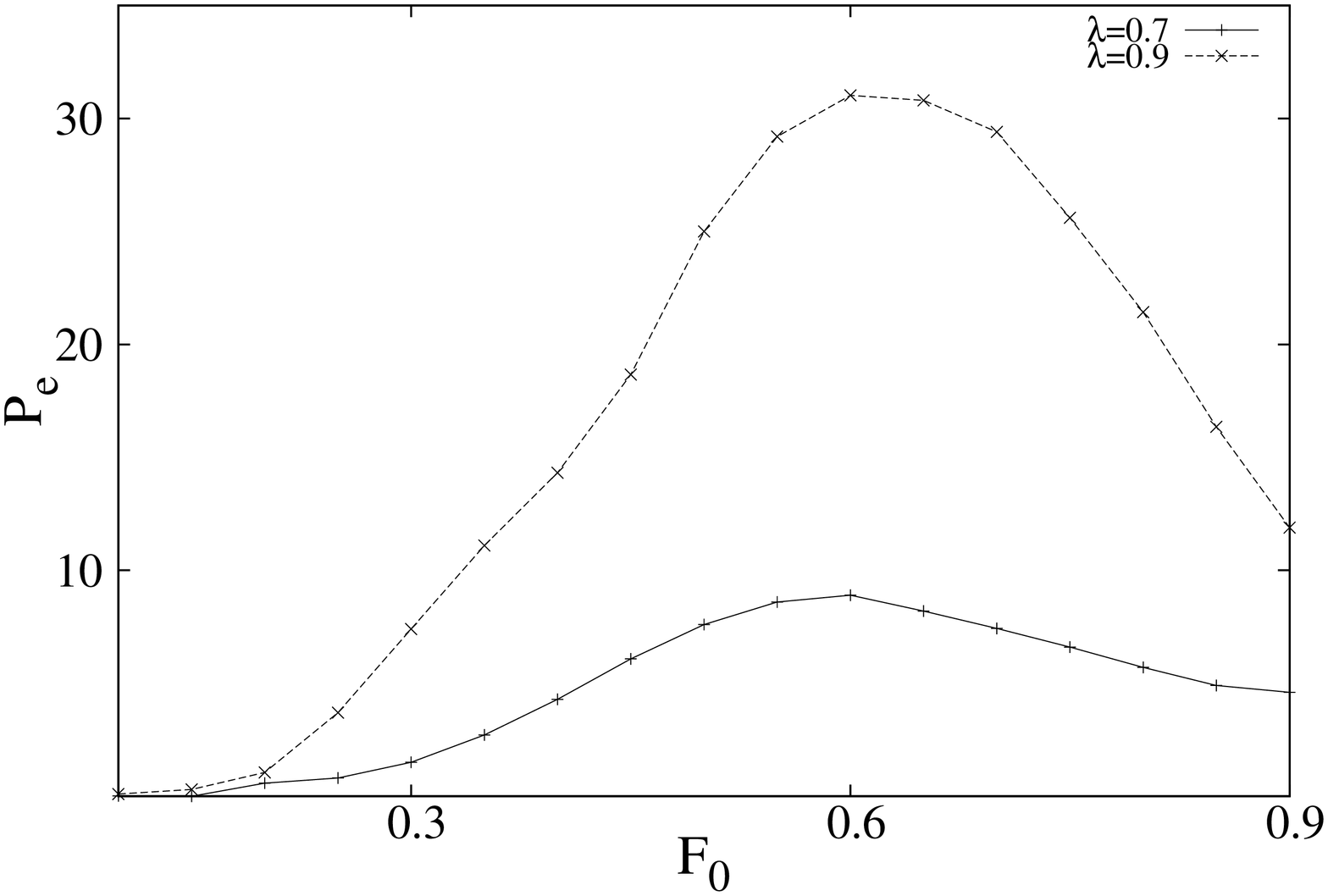}}
\hspace{0.01cm}
\subfigure[]{\includegraphics[width=6cm,height=6.7cm,angle=-90]{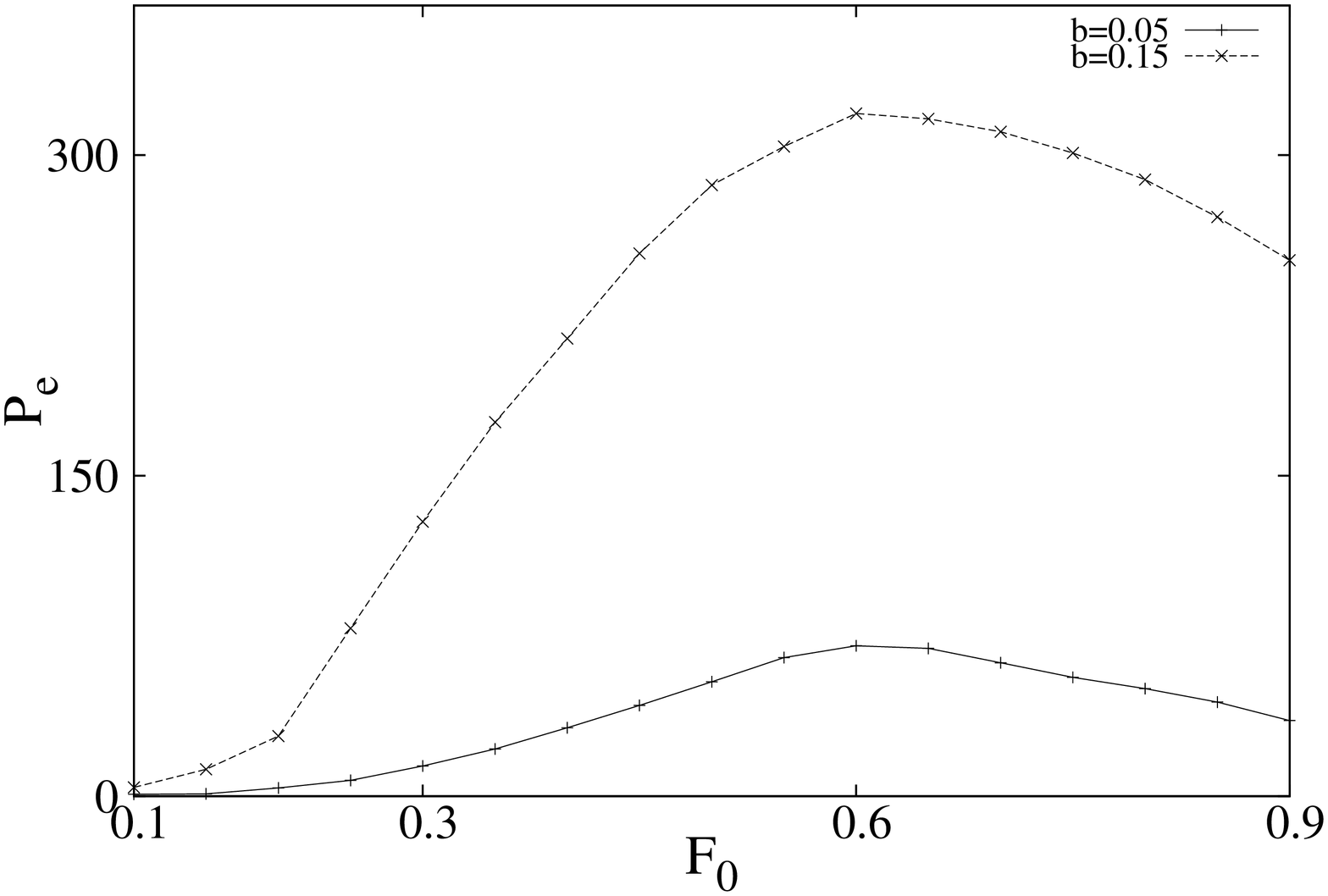}}

\caption{Plot of Peclet number $P_e$ versus $F_0$ for different values of 
$\lambda$ (a) and $b$ (b) for $\gamma_0=0.12$, $\tau =1000$, $\phi=0.35$ and
$T=0.4$.} 
\label{fig:edge}
\end{figure}

\par A measure of the 
average velocity $\langle \bar v \rangle$ or the diffusion coefficient
$D_{eff}$ in isolation will not give the complete picture of the nature of 
transport in a ratchet. For, a ratchet with high average particle velocity can have
high diffusion associated with the transport making the transport incoherent.
On the other hand, even transport with lower $D_{eff}$ does not guarantee higher
average velocity. So it is necessary to measure the coherency of transport
in a ratchet. A parameter which gives a measure of the coherency of 
transport and includes both the average velocity $\langle \bar v \rangle$ 
and the diffusion coefficient $D_{eff}$, is the Peclet number 
$P_e$ \cite{Landauc1}. 

\par Fig. 4a and 4b shows the plot of Peclet number ($P_e$) for the 
{\it{inhomogeneous ratchet}} and the {\it{asymmetric ratchet}} respectively. 
For both the types of 
ratchets, Peclet number is much higher than 2 which proves
that though the particle transport is highly diffusive, it is greatly coherent 
too. However coherency of transport in the {\it{asymmetric ratchet}} is much higher 
than that of the {\it{inhomogeneous ratchet}}. The coherency of transport
is dependent on the degree of asymmetry in the system (i.e. the values of
 $\lambda$ and $b$) in both the types of ratchets.

\par In the ratchet models under our consideration, though there is no external
load, the ratchet does work in transporting particles against the viscous drag
of the medium in which it operates. So it is necessary to have a measure of 
its efficiency.  
 A definition of efficiency which 
takes into account the work done solely against the viscous drag is called the 
{\it Stokes efficiency}, 
$\eta_{S}=\frac{\langle v \rangle ^2}{|\langle v^2 \rangle - T|}$ \cite{MachuraAc1}. 

\begin{figure}[htp]
  \centering
  \subfigure[]{\includegraphics[width=6cm,height=6.7cm,angle=-90]{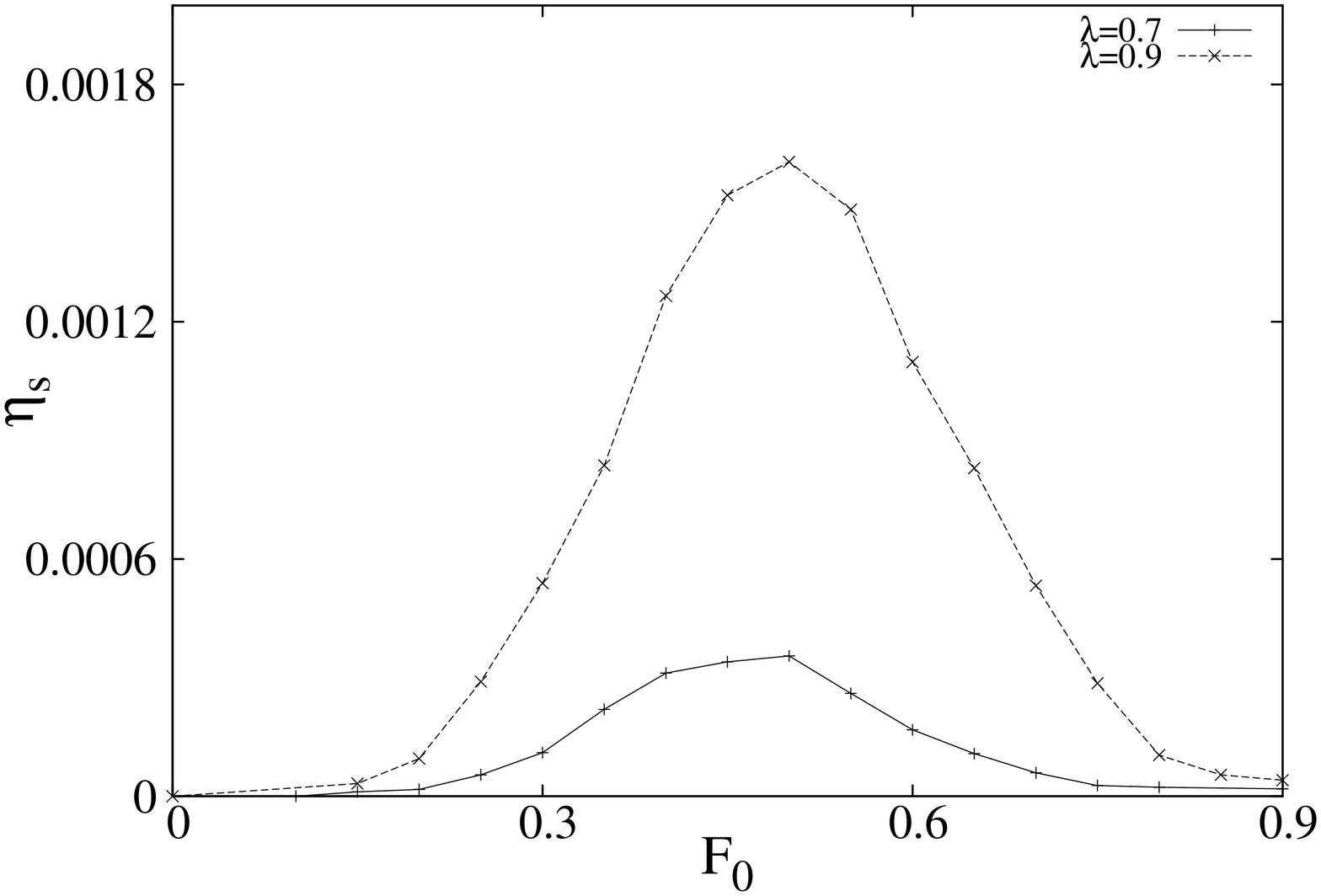}}
\hspace{0.01cm}
\subfigure[]{\includegraphics[width=6cm,height=6.7cm,angle=-90]{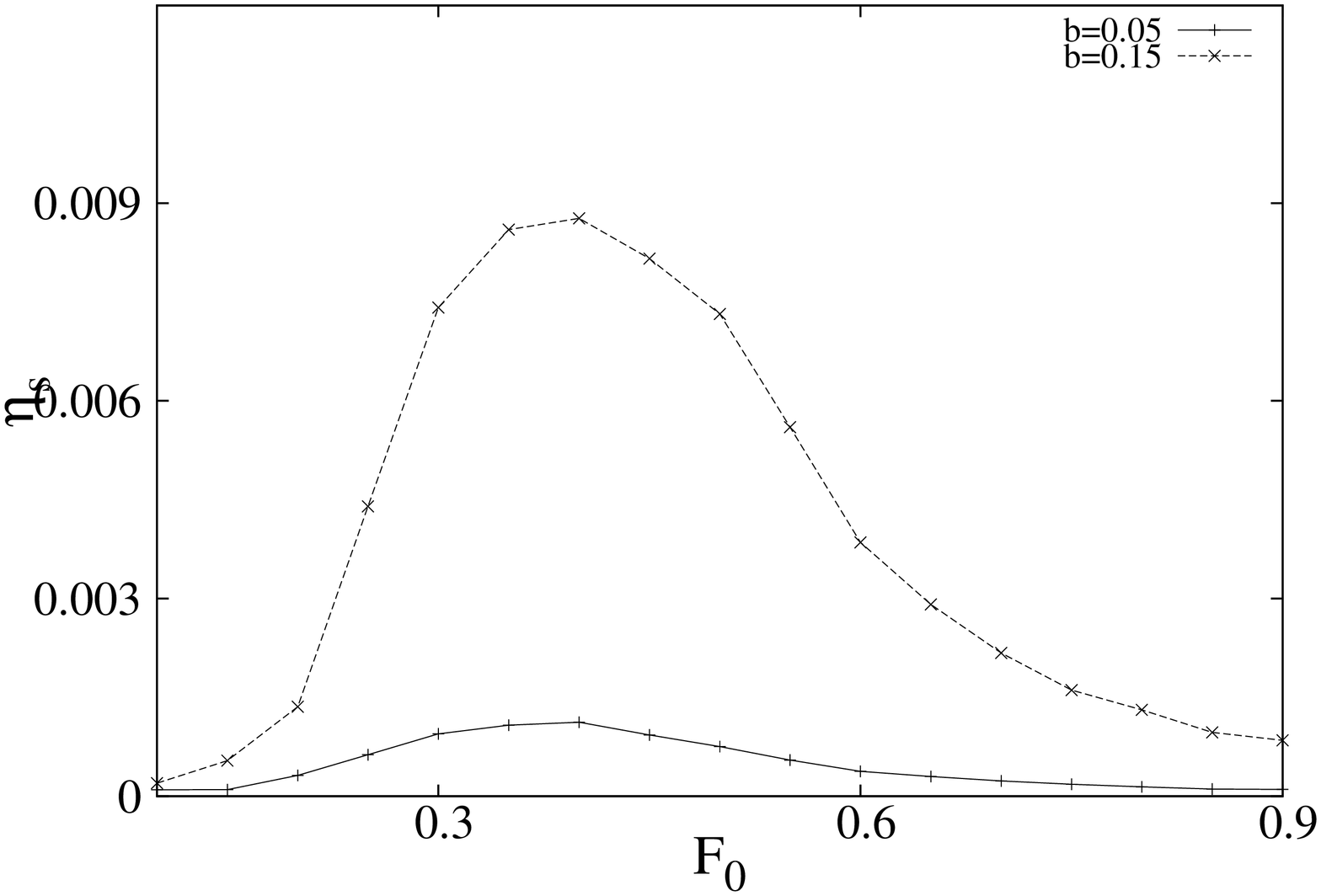}}

\caption{Plot of Stoke's efficiency $\eta_s$ versus $F_0$ for different values of 
$\lambda$ (a) and $b$ (b) for $\gamma_0=0.12$, $\tau =1000$, $\phi=0.35$ and
$T=0.4$.} 
\label{fig:edge}
\end{figure}

Fig. 5 gives a plot of the Stoke's efficiency as a function of the external
driving force $F_0$ for the two kinds of ratchets under our consideration
for the parameters mentioned and for different values of the asymmetry
parameter. The efficiency of performance of the ratchets 
are not only dependent
on the degree and type of asymmetry but also on the amplitude of external 
drive. The efficiency for both the kinds of ratchets peak around the drive
amplitude where there is a peaking of the average particle velocity.

\begin{figure}[htp]
  \centering
  \subfigure[]{\includegraphics[width=6cm,height=6.7cm,angle=-90]{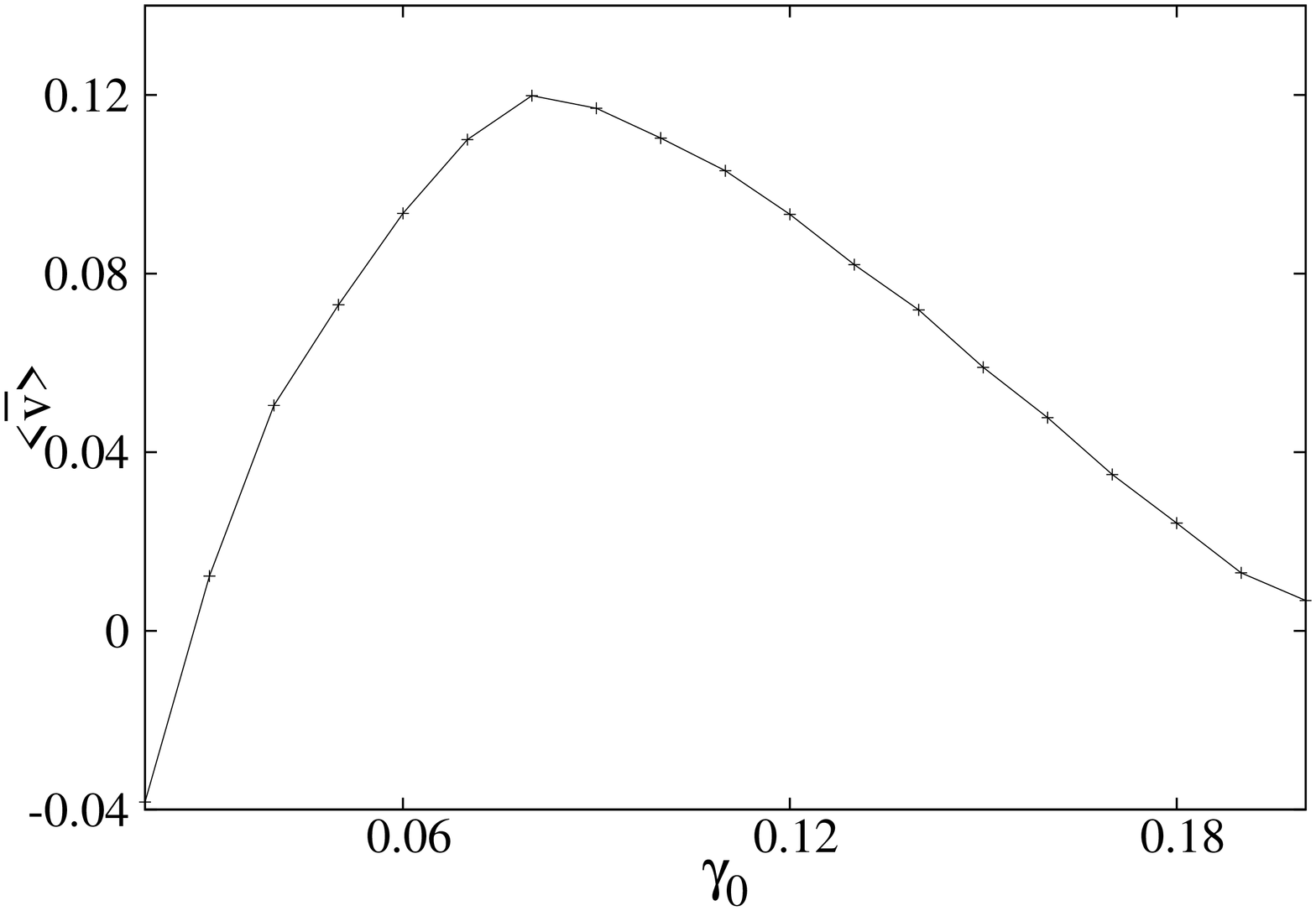}}
\hspace{0.01cm}
\subfigure[]{\includegraphics[width=6cm,height=6.7cm,angle=-90]{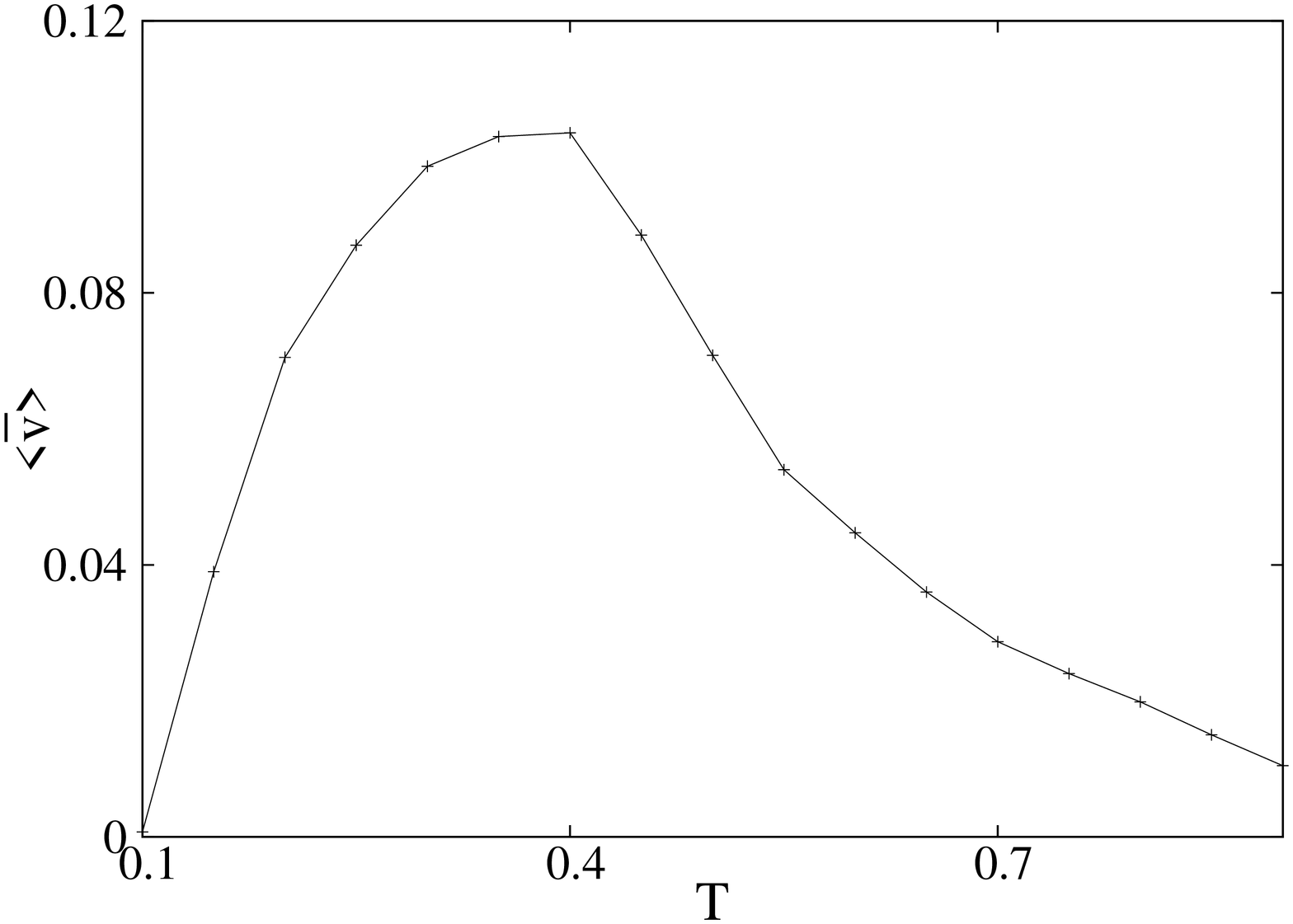}}

\caption{Plot of $\langle \bar v \rangle$ versus friction coefficient $\gamma_0$(a)
 and versus temperature $T$(b) for $\lambda=0.9$, $F_0=0.5$, $\tau =1000$, 
$\phi=0.35$ and $T=0.4$.} 
\label{fig:edge}
\end{figure}

\par We also have studied the inhomogeneous ratchet as a function of the 
other system parameters. Fig. 6a and Fig. 6b shows the plot of 
$\langle \bar v \rangle$ as a function of the friction coefficient 
$\gamma_0$ and temperature $T$ for fixed values of the drive amplitude $F_0$ 
and phase difference $\phi$. 
$\langle \bar v \rangle$ peaks for an optimum value of the friction coefficient
$\gamma_0$ and temperature $T$.

\par When the friction in the system is very low, the cause of asymmetry in the 
inhomogeneous
system is very feeble. As a result of which the difference between the particle 
hoppings to either side of a potential well will not be high which results in 
low particle current. As the friction is increased, the particle sees more 
asymmetry leading to an increase in current with the friction coefficient, 
ofcourse other parameters remaining fixed. However if friction is too high again
particle mobility will reduce leading to less current. This explains the
peaking of particle current at an optimum value of the frction coefficient.
\par The observed variation of particle current with temperature also can
be understood by considering the fact that at low temperatures, the
particles will have less probability of coming out of the potential wells
resluting in less current. As temperature increases, the particles starts 
jumping out of the wells more frequently, and asymmetry in the system
giving rise to a difference in the hoppings to the right and the left
of the well, giving rise to a finite current. At very high temperatures 
though the barrier to particle motion and hence the effect of asymmetry 
almost vanishes reducing the current. In between these two extremes,
expectedly the particle current peaks at an optimum value of temperature.

\section {Effect of both inhomogeneity and asymmetry in potential}
We explore the pssibility of combining features of the 
{\it inhomogeneous ratchet} and the {\it asymmetric ratchet} to design
an underdamped ratchet model having better transport characteristics and 
efficiency of transport. For this, a small asymmetry is added to the 
underlying periodic potential (finite value of $b$) of the 
{\it inhomogeneous ratchet}. Using this {\it combination ratchet}, we than 
calculate the various parameters characterising the transport.

\begin{figure}[htp]
  \centering
  \subfigure[]{\includegraphics[width=6cm,height=6.7cm,angle=-90]{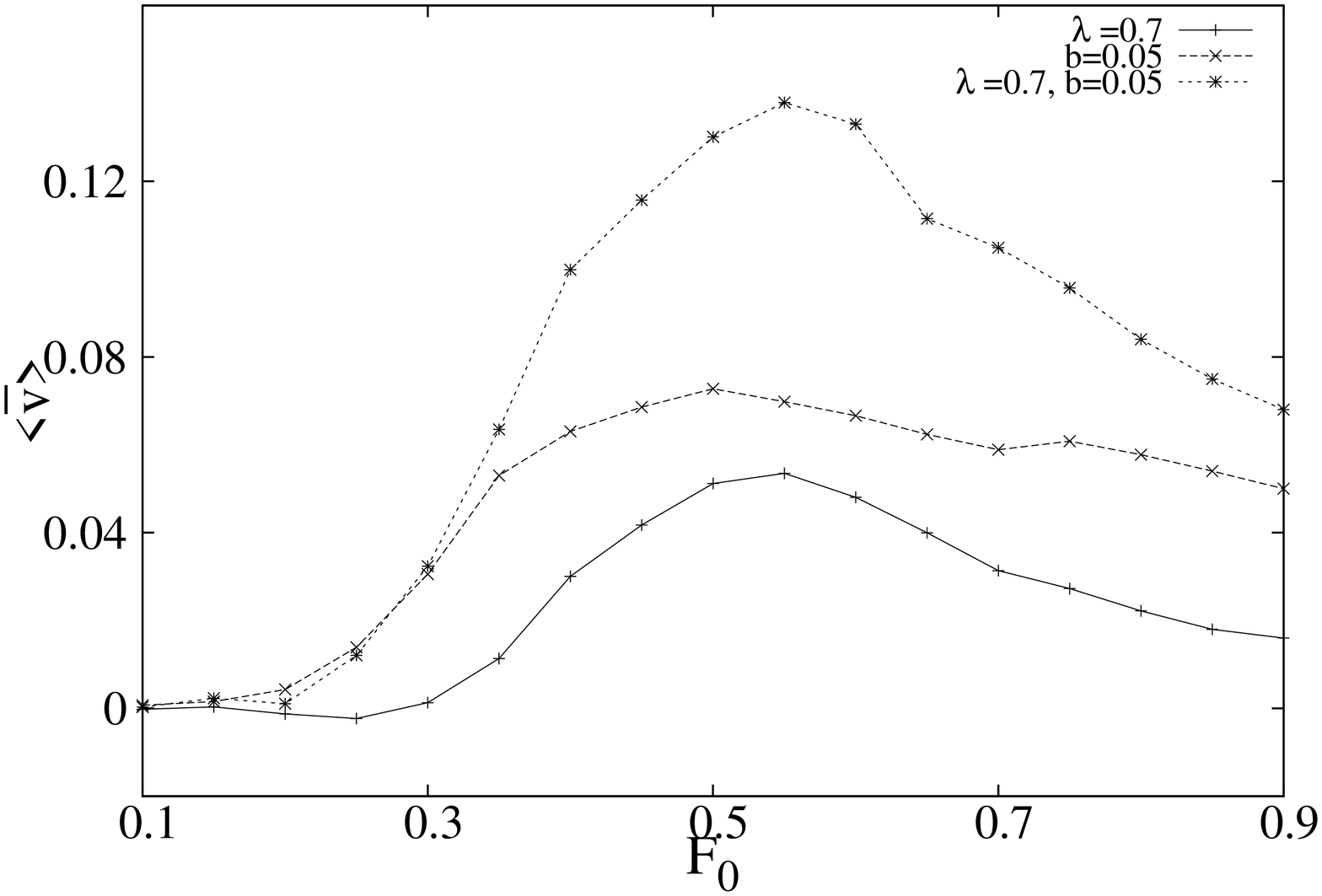}}
\hspace{0.01cm}
\subfigure[]{\includegraphics[width=6cm,height=6.7cm,angle=-90]{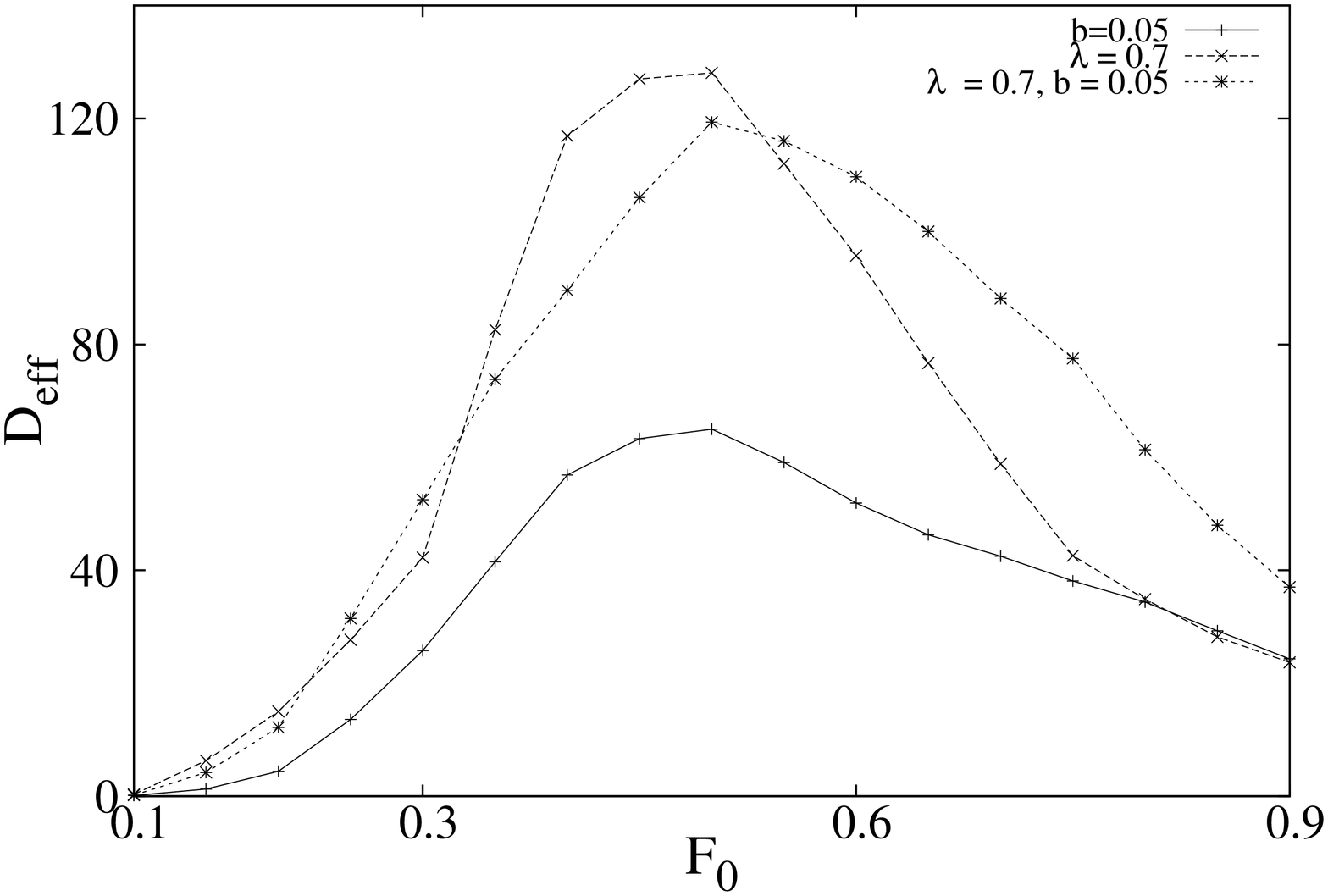}}
\hspace{0.01cm}
  \subfigure[]{\includegraphics[width=6cm,height=6.7cm,angle=-90]{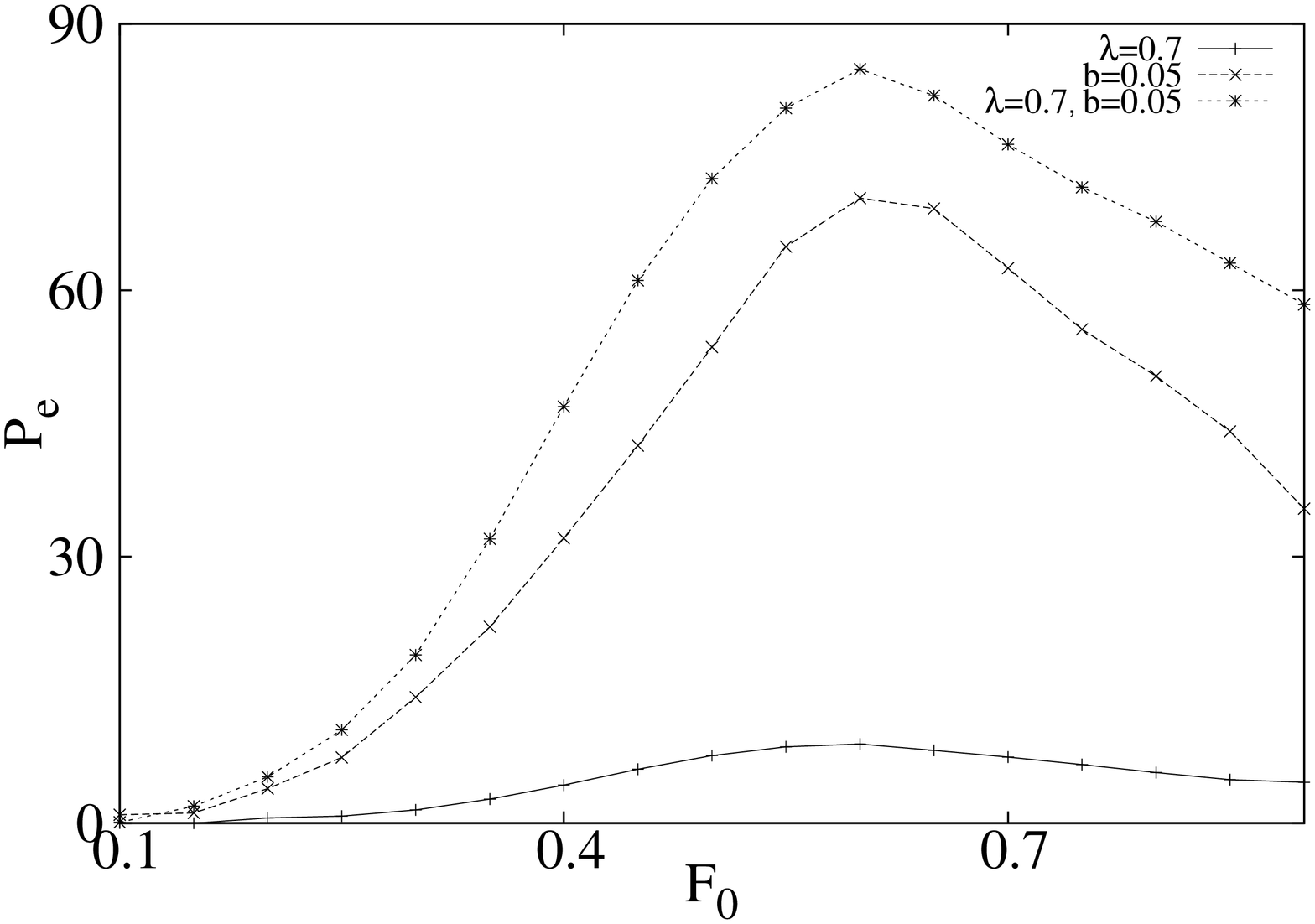}}
\hspace{0.01cm}
\subfigure[]{\includegraphics[width=6cm,height=6.7cm,angle=-90]{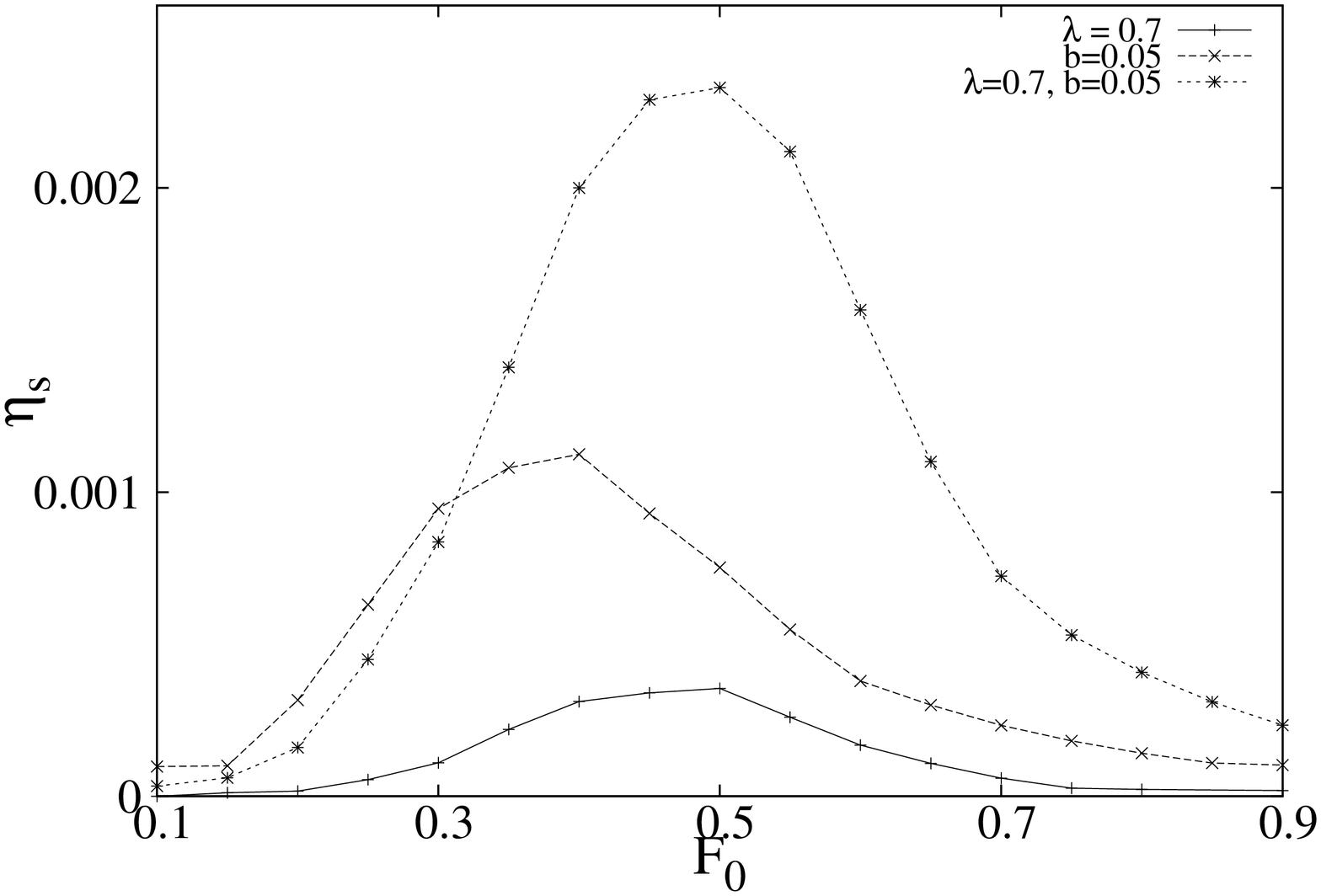}}
\caption{Plot of the different parameters for the {\it combination ratchet}
in comparision to the {\it inhomogeneous ratchet} and the {\it asymmetric 
ratchet} as a function of $F_0$; $\langle \bar v \rangle$ (a), 
$D_{eff}$ (b), $P_e$ (c) and $\eta_s$ (d); $\tau =1000$, $\phi=0.35$ and $T=0.4$.} 
\label{fig:edge}
\end{figure}

\par In Fig. 7 the plots of the various parameters are given for the 
{\it combination ratchet}. It can be clearly
seen that for the particular values considered ($\lambda=0.7, b=0.05$), the 
average particle velocity for the {\it combination ratchet} is substantially 
higher than that for the individual ratchets (Fig. 7a). Interstingly, the value of the 
diffusion coefficient $D_{eff}$ remains almost unchanged (Fig. 7b). Thus with
the {\it combination ratchet}, higher particle current is obtained without
compromising on the quality of transport. This is also reflected
from a measure of the Peclet number (Fig. 7c). Also the efficiency of the
{\it combination ratchet} is comparatively much higher than that for the
individual  
{\it{asymmetric ratchet}} and the {\it{inhomogeneous ratchet}} (Fig. 7d). 

\par The {\it combination ratchet} model, which combines essential features
of the {\it inhomogeneous ratchet} and the {\it asymmetric ratchet} 
therefore provides a new ratchet model in the underdamped regime with 
better transport characteristics and higher efficiency.

\section{Conclusion}
 Study of ratchets has been an active field of research over the past few 
decades particulary because of its practical utility in explaining different
physical and biological systems. Different models of ratchets have been 
proposed and studied. We have studied underdamped particle transport in 
periodic potentials. This finds many analogies in naturally occuring systems as
already discussed in the introduction. 
Primary focus was to study transport in an {\it{inhomogeneous ratchet}} with 
a symmetric and periodic potential but with a space dependent friction coefficient.
The performance
of the inhomogeneous ratchet as a function of the different parameters is studied.
The transport characteristics are dependent on the various parameters governing
the system. 
The performance of this ratchet is compared with that of the 
{\it{asymmetric ratchet}} with an asymmetric periodic potential but a homogeneous
friction coefficient. It is observed that though inhomogeneity is a feeble
cause of asymmetry in the system, substantial ratchet current can be obtained
with the model {\it{inhomogeneous ratchet}}. 
\par We propose a new scheme of a ratchet
model combining both inhomogeneity in the system and asymmetry in the underlying
potential. This {\it combination ratchet} model has 
comparatively much better performance characteristics in the underdamped regime.  
\par Also, the performance characteristics of the ratchet model being sensitively
dependent on the choice of the various parameters governing the system, a careful 
choice of the different parameters becomes a necessity 
to optimise the working of the ratchet.

{\bf{ACKNOWLEDGEMENTS}}

The author acknowledges financial support received from University Grants Commission (UGC),
Government of India under Project No. F.5-152/2012-13/MRP/NERO/565 for carrying out this work

\end{document}